# First-Principles Thermodynamics of Graphene Growth on Cu Surface

*Wenhua Zhang, Ping Wu, Zhenyu Li,\* Jinlong Yang.*

Hefei National Laboratory for Physical Sciences at Microscale, University of Science and Technology of China, Hefei 230026, China

\*To whom correspondence should be addressed. E-mail: zyli@ustc.edu.cn

**ABSTRACT:** Chemical vapor deposition (CVD) is an important method to synthesis grapheme on a substract. Recently, Cu becomes the most popular CVD substrate for graphene growth. Here, we combine electronic structure calculation, molecular dynamics simulation, and thermodynamics analysis to study the graphene growth process on Cu surface. As a fundamentally important but previously overlooked fact, we find that carbon atoms are thermodynamically unfavorable on Cu surface under typical experimental conditions. The active species for graphene growth should thus mainly be $CH_x$ instead of atomic carbon. Based on this new picture, the nucleation behavior can be understood, which explains many experimental observations and also provides us a guide to improve graphene sample quality.

**Introduction**

A method to massively produce high quality samples is crucial for the graphene research. Recently, chemical vapor deposition (CVD) has been widely used in graphene synthesis, [1], [2], [3] with many different metals used as the substrate. Among them, Ni has good lattice match with graphene, while Cu is superior for controllably growing single layer graphene.[2] An isotope-effect study indicates that, during CVD growth of graphene on Ni substrate, carbon atoms dissolve into the bulk. Contrarily, the growth process is limited on the surface in the Cu case, [4] which makes Cu very attractive for graphene growth. Recently, an intense experimental effort has been devoted to graphene growth on Cu, [5],[6],[7],[8],[9],[10],[11],[12] and it turns out that the sample quality is sensitive to various experimental conditions. Different recipes have been proposed to obtain high quality graphene. For example, a two-step process with low pressure first and then higher pressure is used to produce large single-domain graphene.[7]

To optimize graphene growth conditions, it is highly desirable to understand the underlying atomic details of growth. Unfortunately, at the atomic scale, our knowledge about graphene growth on Cu is still very limited. For example, even the most fundamental question, what the active surface species is at the initial stage of growth, remains unclear. Based on experiences from other CVD substrates, atomic carbon has been supposed to be the answer, [4],[13] without taking the uniqueness of Cu into account. In this article, hydrocarbon decomposition and graphene nucleation on Cu surface are studied from first principles, and many insights into the mechanism of graphene growth are obtained.

**Computational Details**

Electronic structure calculations are performed with the DMol$^3$ implementation [14],[15] of density functional theory (DFT) using the PBE exchange-correlation functional.[16] Double-zeta numerical basis set with polarization functions (DNP) and DFT semi-core pseudopotential (DSPP) are used for all atoms. A thermal smearing of 0.002 Hartree and a real-space cutoff of 4.4 Å are adopted. In calculations of dehydrogenation reactions, Cu (111) and (100) surfaces are simulated with a five-layer $p(3\times3)$ slab with a ~15 Å vacuum. During geometry optimizations, all atoms are relaxed, except those in the bottom two layers, which are kept at their bulk positions. A 4×4×1 k-point grid is adopted. All the calculations are performed within the spin-polarized frame. To calculate the carbon chemical potential for large graphene clusters ($C_{24}$ and $C_{54}$) adsorbed on Cu surfaces, a $p(6\times6)$ supercell is used. The transition state search is performed with the synchronous transit methods.[17]

First-principles molecular dynamics (MD) simulations are performed using the VASP code[18] for CH covered Cu (111) and (100) surfaces, with a three-layer $p(6\times6)$ surface model. The CH coverage is set to 0.33 ML on the Cu (111) surface and 0.42 ML on the Cu (100) surface. Gamma-only $k$-point sampling is used. MD initial configuration is obtained from a geometry optimization with the bottom layer fixed to its Cu bulk atomic positions. For each system, canonical (NVT) ensemble MD simulation is performed at 1300 K for 4.0 ps with a 1.0 fs time step. Temperature reaches equilibrium after 0.5 ps. Therefore, the production trajectory lasts for 3.5 ps.

**Results and Discussion**

**A. Dehydrogenation energetics**

First, we consider the dehydrogenation of $CH_4$ on Cu surfaces. The initial state is an adsorbed $CH_4$ molecule, and the final product is a C atom plus four H atoms on the surface. There are three intermediates, namely methyl ($CH_3$), methylene ($CH_2$), and methylidyne (CH), corresponding to the following four elementary steps:

$$CH_4 \rightarrow CH_3 + H$$

$$CH_3 \rightarrow CH_2 + H$$

$$CH_2 \rightarrow CH + H$$

$$CH \rightarrow C + H$$

Geometric structures of the initial, transition, and final states for each elementary dehydrogenation step on Cu (111) surface are shown in Figure 1. In our model, the adsorption geometries of intermediate species as the final state of a dehydrogenation step and as the initial state of the next dehydrogenation step can be slightly different, due to the removal of the dissociated H in the latter step. Our test calculations show that the energy difference caused by such a geometry difference is small. Reaction energies, activation energies, and some key geometric parameters are listed in Table 1. As also shown in Figure 2, all four dehydrogenation steps are endothermic, and the corresponding activation energy barriers are about 1.0-2.0 eV. The final product C+4H is already 3.60 eV higher in energy than the adsorbed $CH_4$, which suggests that atomic carbon is energetically very unfavorable on Cu surface.[19]

In most experiments, copper foil is used instead of single crystal. Therefore, different Cu surfaces can be exposed. We thus also investigate the dehydrogenation processes on Cu (100) surface. Geometric structures are shown in Figure 3, and some energetic and geometric data are listed in Table 2. Both hollow and bridge sites are obtained for the dissociated H in different dehydrogenation steps, while the energy difference caused by these two different H adsorption sites is as low as 0.06 eV. Atomic carbon on Cu (100) surface has an adsorption energy of 6.08 eV. It is 1.23 eV more stable than on the Cu (111) surface. This can be easily understood, since the Cu coordination number of C is higher on the (100) surface. Despite the C adsorption energy difference, the whole energy profile of the dehydrogenation process is still very similar for these two surfaces. There is also a large total energy increase (2.75 eV) for methane dehydrogenation on the (100) surface. Therefore, the dehydrogenation energetics is not expected to be very sensitive to surface morphology. We have also tested different carbon sources. For example, similar endothermic behavior is obtained for ethylene ($C_2H_4$) on Cu (111) surface. During the four dehydrogenation steps, its energy increases 0.48, 0.09, 0.67, and 0.62 eV, respectively.

**B. Alternative reaction pathway**

Atomic carbon is commonly used as the starting point to study graphene growth on metal surfaces.[4],[13],[20],[21] On active metal surfaces, such as Pd and Ru, decomposition of $CH_4$ is exothermic.[22],[23] Therefore, atomic carbon is readily available. On Ni surface, although decomposition of $CH_4$ is slightly endothermic (with the final C atom product in a subsurface site),[24],[25] atomic carbon is kinetically very favorable due to the relatively high solubility of C in Ni. Therefore, on these surfaces, atomic carbon is expected to make a central role in graphene growth. While on Cu surface, according to our calculations, dehydrogenation of hydrocarbon is very difficult to complete, which strongly suggests alternative reaction paths. It is more likely partially dehydrogenated species, such as $CH_x$, will combine with each other before going to the final hydrogen-free product on the Cu surface. Of course, to grow graphene, dehydrogenation should finally be completed. We note that it can be completed at a very late stage with large $C_xH_y$ structures already formed, as demonstrated by a recent experiment.[26]

To validate this picture, we perform MD simulations on the CH covered Cu (111) surface at 1300 K. During the 3.5 ps trajectory, we see very significant surface structure relaxations (Figure 4a), which is a natural result of the high simulation temperature (already very close to the melting point of Cu). As shown in Figure 4c, no CH dissociation is observed, except occasional oscillations of H atoms between their neighboring C and Cu atoms. This can be understood by the dehydrogenation energetics. Although the CH dissociation barrier has a chance to be conquered at a temperature as high as 1300 K, there is a large thermodynamic driving force to recombine C and H. As a result, complete dissociation is not observed in the time scale of our simulation.

Most importantly, $C_2H_2$ can be easily formed on Cu (111) surface, which represents a more favorable reaction path compared to CH dissociation. Consistently, a transition state calculation for CH combination on Cu (111) surface predicts an exothermic reaction energy of 1.94 eV and an activation barrier only 0.3 eV. Further combination of $C_2H_2$ with an additional CH is also exothermic (0.98 eV) with a higher energy barrier about 1.1 eV. This relatively high barrier for the $C_2H_2$+CH reaction is due to the formation of a C-Cu-C bridge structure (Figure S3) when CH approaching $C_2H_2$. Such a kind of bridging metal (BM) structures has been discussed previously,[13] when considering coalescence of carbon atoms and clusters. Now, we can see that the BM structure is quite universal on the soft Cu surface, with both carbon and hydrocarbon species adsorbed.

On Cu (100) surface, the Cu lattice distortion during the MD simulation is much smaller than the (111) surface (Figure 4b). A possible reason is that, with a strong Cu-C interaction compared to Cu-Cu interaction,[13] Cu atoms on the denser (111) surface are easier to be pulled out by C contained species. Another observation is that the diffusion of CH groups is much slower on the (100) surface. During the whole 3.5 ps trajectory, we only see a single CH combination event in the last ps, and the formed $C_2H_2$ then desorbs from the surface. This desorption is a result of the relatively smaller distortion of the (100) surface, since an ideal surface provides less binding to acetylene. Similar to the (111) surface, no CH dissociation is observed on this surface. The difference is that the possibility to observed H oscillation between neighboring C and Cu is higher on the (100) surface.

**C. Nucleation Thermodynamics**

In the last section, we have demonstrated that there are possibly more favorable reaction paths to grow graphene compared to a complete dehydrogenation of $CH_4$ first. Therefore, at an early stage of graphene growth, active species are expected to contain hydrogen. A possible picture is: small carbon contained species are not stable without H. Only when they grow to a certain size, they become stable without H. At that point, we obtain the smallest graphene. With this picture, an interesting question is when such a hydrogen-free graphene flake can be nucleated. We will answer this question by a thermodynamic analysis.

We consider a typical experimental growth temperature $T=1300$ K. With the reference pressure $P_0=1$ bar, and taking one-half of the energy of $H_2$ molecule as a reference, according to the NIST-JANAF thermochemical tables,[27] we have the following expression on chemical potential of hydrogen in the unit of eV

$$\mu_H(T,P_0) = \frac{1}{2}[h(T,P_0) - h(0,P_0) - TS(T,P_0)] = -0.975$$

Then, under arbitrary pressure, we have

$$\mu_H(T,P) = \mu_H(T,P_0) + \frac{1}{2}k_B T \ln \frac{P_{H_2}}{P_0} = -0.975 + 0.056 \ln \frac{P_{H_2}}{P_0}$$

For methane, we have

$$\Delta g_{CH_4}(T,P_0) = h(T,P_0) - h(0,P_0) - TS(T,P_0) = -2.868$$

Taking one-half of the $H_2$ energy and the C atom energy as a reference, the DFT-calculated energy of $CH_4$ is -9.234 eV. So, we have

$$g_{CH_4}(T,P) = -12.102 + 0.112\ln\frac{P_{CH_4}}{P_0}$$

In a typical experiment, $CH_4$ and $H_2$ are fed in simultaneously. Suppose the ratio of the partial pressures of $CH_4$ and $H_2$ is $\chi$, then we have the following relationship between the chemical potentials of C and H at the equilibrium of $CH_4$ and $H_2$

$$\mu_C = g_{CH_4} - 4\mu_H$$
$$= -2\mu_H - 10.152 + 0.112\ln\chi$$

$\mu_H$ is readily related to the experimental partial pressure of $H_2$ within the ideal gas approximation. Therefore, at a fixed $\chi$, $\mu_C$ during CVD growth is also related to the $H_2$ partial pressure, as shown in Figure 5. When a surface carbon species has a chemical potential higher than $\mu_C$ (i.e. not in the yellow area), it is not stable and will react with $H_2$.

Based on this thermodynamic model, we can consider the stability of atomic carbon adsorbed on Cu (111) surface. The chemical potential of an isolated atomic carbon on the surface can be approximated by its adsorption energy (-4.85 eV, the horizontal line marked with $C_1$ in Figure 5). Under most experimentally accessible pressures, it is much higher than $\mu_C$. Therefore, atomic carbon is not stable, consistent with our energetics results. We have also checked the Cu (100) surface and two step sites, atomic carbon there is also unstable at typical experimental conditions. Carbon chemical potential is -6.08 for atomic carbon on the (100) surface, and it is -5.78 and -5.74 eV at the (211) and (411) steps, respectively.

The carbon chemical potential of two-dimensional graphene adsorbed on Cu surfaces is calculated by adding an adsorption energy correction to the chemical potential of a free graphene (-7.85 eV). In adsorption energy calculations, the lattice parameter of graphene is expanded to accommodate that of Cu (Figure 6). The calculated adsorption energy is 0.01 eV/C for both (111) and (100) surfaces. It is well known that GGA underestimates van der Waals interactions.[28] However, even with the underestimated adsorption energy, carbon chemical potential of graphene (the lowest horizontal line in Figure 3) is still lower than that of the source gases in most cases. This is the thermodynamic driving force to grow graphene. We note that graphene can also become unstable under very high pressures, where it is expected to be etched by $H_2$ and form $CH_4$.

Infinite two-dimensional graphene is stable on the surface, while the smallest carbon cluster, atomic carbon, is not stable. Then, a natural question is when the transition from unstable small carbon clusters to stable graphene flakes happens? The answer of this question gives a rough estimation on the nucleation size of graphene growth. To answer it, we consider the potential energy ($\mu_C^n$) of a carbon cluster with $n$ atoms adsorbed on the Cu surface, which is approximated as:

$$\mu_C^n = \frac{1}{n}(E_{C_n/Surf} - E_{Surf}) - E_C$$

where $E_{C_n/Surf}$ is the energy of the adsorbed system, and $E_{Surf}$ is the energy of the Cu surface, and $E_C$ is the energy of an isolated carbon atom in vacuum. The $n=1$ atomic carbon case we have just discussed. Besides that, we also calculate $\mu_C^n$ for $C_6$, $C_{24}$, and $C_{54}$ clusters adsorbed on both Cu (111) and (100) surfaces. The obtained potential energies are listed in Table 3, and the optimized geometries of carbon clusters are shown in Figure 6.

With the potential energy of different carbon clusters at hand, we can estimate the nucleation size. For example, under a $H_2$ partial pressure of $10^{-5}$ bar and a $CH_4$ partial pressure of $2\times 10^{-4}$ bar, the nucleation size is estimated to be about 6 carbon atoms on (111) surface as read from Figure 5a. However, if the $CH_4/H_2$ ratio decreases, the nucleation size can be increased to as large as about 24 for $\chi=1/20$ (the red line in Figure 5a). Similar analysis can be made for ethylene, another frequently used carbon source in graphene growth. For ethylene, we have

$$\mu_C = -\mu_H - 8.692 + 0.056 \ln \chi$$

An interesting difference to $CH_4$ is that, at a fixed $\chi$, the carbon chemical potential $\mu_C$ is less sensitive to the hydrogen partial pressure (Figure 5b).

**Discussion and conclusions**

We emphasize that the identification of active surface species is of fundamental importance for graphene growth. On Cu surface, unlike many other metal surfaces, small carbon clusters are more stable than carbon atom.[13],[20] Therefore, if it started from atomic carbon, the growth process would proceed without a nucleation step, which is not consistent with the complicated growth behaviors observed experimentally. Low solubility of C in Cu is a well recognized reason for the surface-limited graphene growth observed in experiment.[4] According to our results, low availability of atomic carbon on the surface is the more direct reason for that.

First-principles thermodynamics based nucleation size estimation provides many insights into graphene growth on Cu. In fact, controlling nucleation size is an important means to improve graphene sample quality and/or productivity. The experimental observation of growth rate increase with the decrease of $H_2$ concentration [5] is a natural result of the smaller nucleation size at higher $CH_4/H_2$ ratio. Lower $CH_4/H_2$ ratio leads to lower nucleation density, and higher quality of single layer sample.[6] Therefore, at the initial stage of graphene growth, if the total pressure is fixed, it is important to keep the partial pressure of $CH_4$ low.[7]

In summary, thermodynamics of graphene growth on Cu surface has been studied. Atomic carbon is found to be not stable on Cu surface, and it is thus not an important species during graphene growth. Based on thermodynamic stability considerations, nucleation size at different experimental conditions can be estimated. Since nucleation size is closely related to the graphene sample quality, it is very desirable to be manually controlled.

**ACKNOWLEDGMENT** We thank Prof. Rodney S. Ruoff for helpful discussions. This work is partially supported by NSFC (20933006, 20803071), by CUSF, by the National Key Basic Research Program (2011CB921404), and by USTC-SCC, SCCAS, and Shanghai Supercomputer Center.

**Supporting Information Available:** $C_2H_4$ dehydrogenation on Cu(111) surface, combination of CH groups, C adsorption at steps, and some test calculations. This material is available free of charge via the Internet at http://pubs.acs.org.

**Table 1.** Relative energies (E) of the initial state (I.S.), transition state (T.S.), and final state (F.S.), distances between detached H and the nearest C ($d_{C-H}$) and Cu ($d_{Cu-H}$) for each elementary step of $CH_4$ dissociation on Cu (111) surface. Numbers in parenthesis refer to energies with the detached H moved far away.

|  | States | E(eV) | $d_{C-H}$(Å) | $d_{Cu-H}$(Å) |
|---|---|---|---|---|
|  | I.S. | 0.00 | 1.081 | 2.397 |
| $CH_4$ | T.S. | 1.77 | 1.980 | 1.535 |
|  | F.S. | 0.89(0.79) | 2.990 | 1.727 |
|  | I.S. | 0.00 | 1.085 | 2.174 |
| $CH_3$ | T.S. | 1.53 | 1.988 | 1.500 |
|  | F.S. | 0.95(0.85) | 2.856 | 1.742 |
|  | I.S. | 0.00 | 1.087 | 2.140 |
| $CH_2$ | T.S. | 1.13 | 1.908 | 1.498 |
|  | F.S. | 0.63(0.58) | 2.873 | 1.739 |
|  | I.S. | 0.00 | 1.077 | 2.685 |
| CH | T.S | 1.98 | 1.787 | 1.572 |
|  | F.S. | 1.49(1.48) | 2.961 | 1.728 |

**Table 2.** Relative energies (E) of the initial state (I.S.), transition state (T.S.), and final state (F.S.), distances between detached H and the nearest C ($d_{C-H}$) and Cu ($d_{Cu-H}$) for each elementary step of $CH_4$ dissociation on Cu (100) surface. Numbers in parenthesis refer to energies with the detached H moved far away.

|        | States | E(eV)       | $d_{C-H}$(Å) | $d_{Cu-H}$(Å) |
|--------|--------|-------------|--------------|---------------|
|        | I.S.   | 0.00        | 1.100        | 2.772         |
| $CH_4$ | T.S.   | 1.59        | 1.792        | 1.591         |
|        | F.S.   | 0.87(0.90)  | 1.921        | 3.371         |
|        | I.S.   | 0.00        | 1.095        | 1.999         |
| $CH_3$ | T.S.   | 1.55        | 2.175        | 1.499         |
|        | F.S.   | 1.08(0.84)  | 1.797        | 3.748         |
|        | I.S.   | 0.00        | 1.109        | 2.012         |
| $CH_2$ | T.S.   | 0.77        | 1.860        | 1.642         |
|        | F.S.   | 0.27(0.12)  | 3.840        | 1.613         |
|        | I.S.   | 0.00        | 1.089        | 2.605         |
| CH     | T.S    | 1.54        | 1.799        | 1.657         |
|        | F.S.   | 0.87(0.89)  | 3.866        | 1.610         |

**Table 3.** Potential energy (eV) of carbon species on Cu (111) and (100) surfaces.

| $\mu_C^n$ (eV) | $C_1$ | $C_6$ | $C_{24}$ | $C_{54}$ | Gr |
|---|---|---|---|---|---|
| Cu(111) | -4.85 | -6.49 | -7.18 | -7.44 | -7.86 |
| Cu(100) | -6.08 | -6.65 | -7.11 | -7.38 | -7.86 |

**Figure 1.** Geometric structures of the initial state (I.S.), transition state (T.S.), and final state (F.S.) of the four steps of $CH_4$ dehydrogenation on Cu (111) surface. Red, white, and gray spheres represent Cu, H, and C atoms, respectively.

**Figure 2.** Energy profile of the dehydrogenation processes of $CH_4$ on Cu (111) and (100) surfaces.

**Figure 3.** Geometric structures of the initial state (I.S.), transition state (T.S.), and final state (F.S.) of the four steps of $CH_4$ dehydrogenation on Cu (100) surface. Red, white, and gray spheres represent Cu, H, and C atoms, respectively.

**Figure 4.** (a) a typical snapshot from MD simulation of a CH covered Cu (111) surface with three $C_2H_2$ groups marked by green circles. (b) a typical snapshot from MD simulation of a CH covered Cu (100) surface. Cu is in red, C in grey, and H in white. Bond length variation of a randomly chosen C-H bond in the MD trajectories for (c) Cu (111) and (d) Cu (100) surfaces.

**Figure 5.** Relationship between $\mu_C$ and partial pressure of $H_2$ during CVD growth of graphene on Cu (111) surface at 1300 K with (a) $CH_4$ or (b) $C_2H_4$ as the carbon source. Black line for $\chi=1$, blue line for $\chi=20$, and red line for $\chi=1/20$. Chemical potentials of carbon clusters adsorbed on Cu (111) surface and that of graphene (Gr) are marked as horizontal lines. Yellow areas mark the stable zones with $\chi=1$. A typical experimental $H_2$ partial pressure ($10^{-5}$ bar) is marked by a vertical line.

**Figure 6.** Geometric structures of carbon cluster and graphene adsorbed on Cu (111) and (100) surfaces. For graphene adsorbed model, the unit cell used in calculations are marked with green dashed lines.

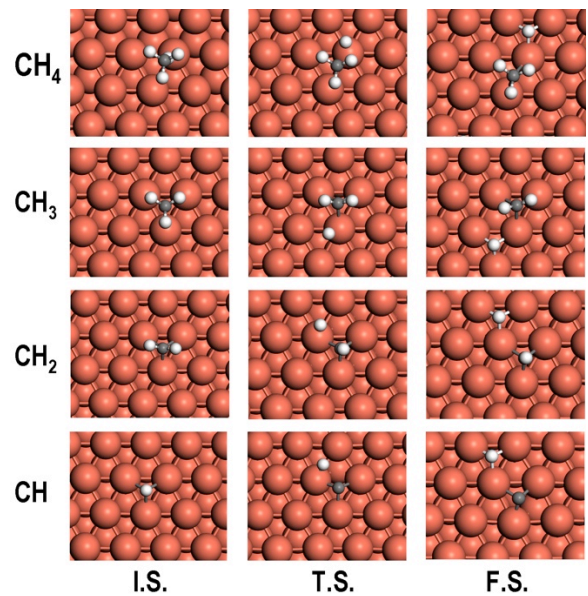

Figure 1. Zhang et al.

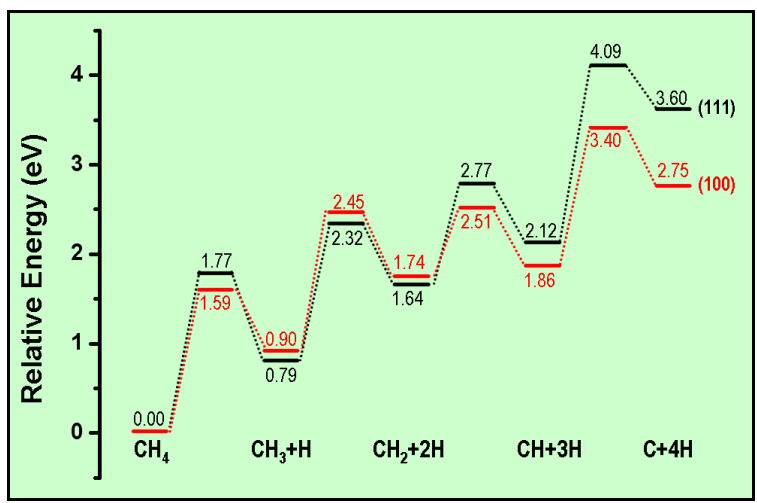

Figure 2, Zhang et al.

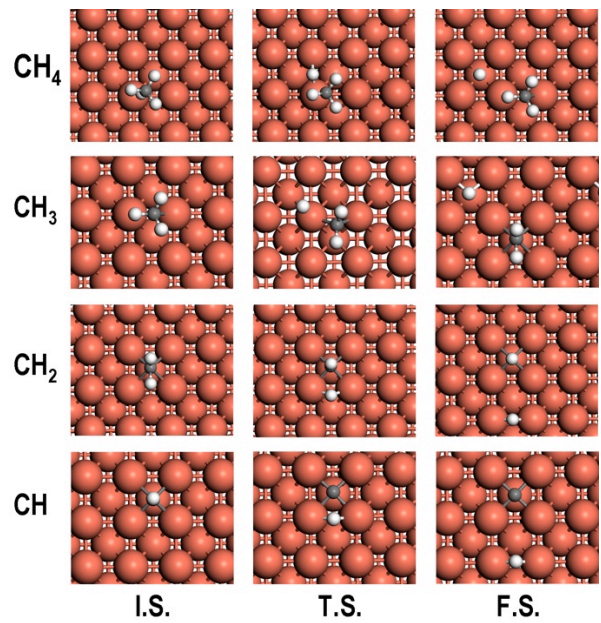

Figure 3, Zhang et al.

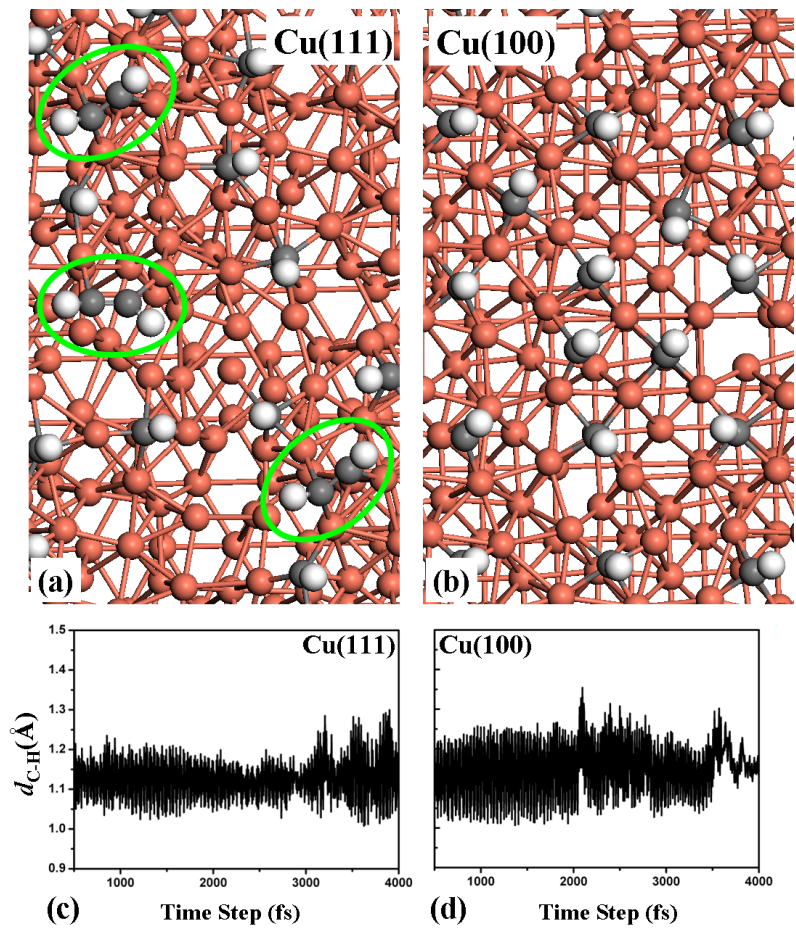

Figure 4, Zhang et al.

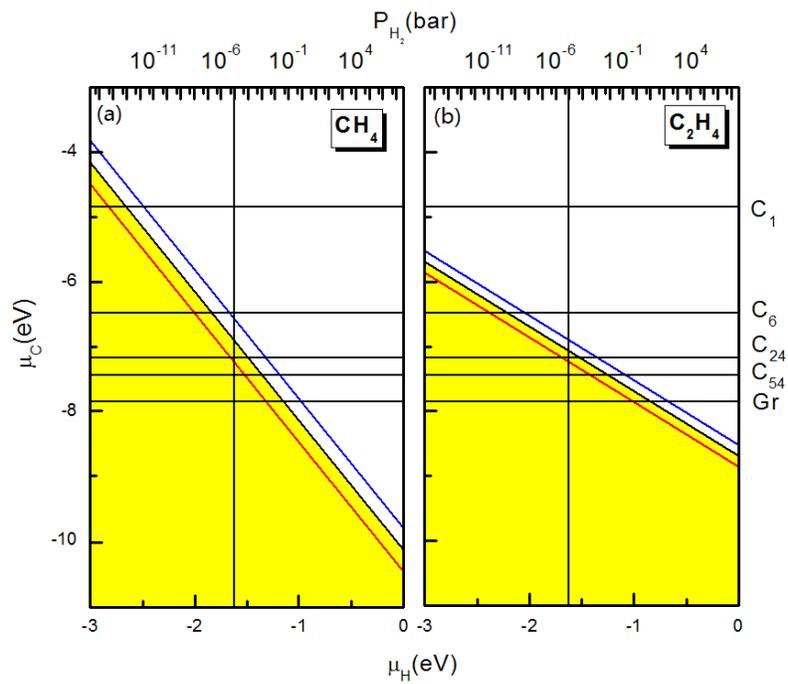

Figure 5, Zhang et al.

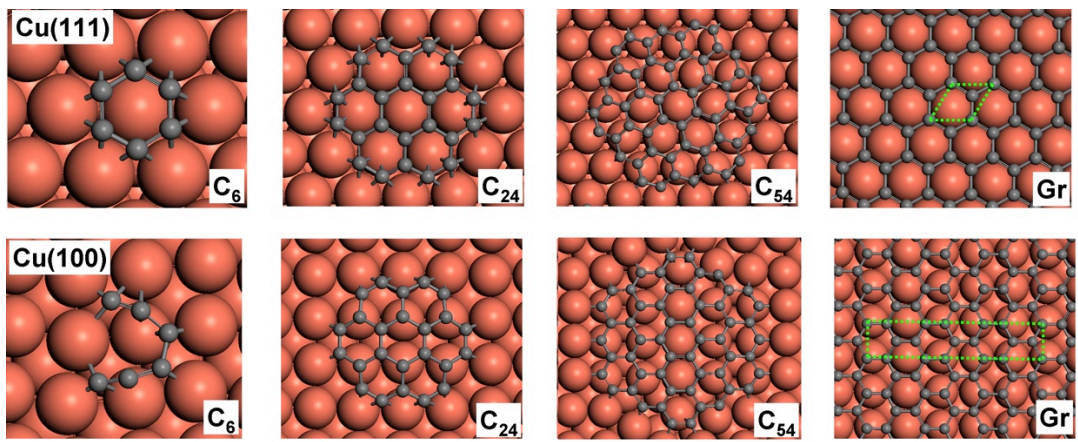

Figure 6, Zhang et al.